\documentstyle[preprint,aps,emlines]{revtex}
\begin{document}
\draft
\title{ Quantization of electromagnetic field in inhomogeneous
dispersive dielectrics}

\author{Z. Hradil}
\address{Department of  Optics,
Palack\'y University\\
17. listopadu 50, 772 07 Olomouc\\ Czech Republic}
\date{}
\maketitle
\begin{abstract}
Canonical quantization  of electromagnetic field inside the
time--spatially  dispersive inhomogeneous dielectrics  is presented.
Interacting  electromagnetic  and matter excitation  fields create the closed
system, Hamiltonian of which may be diagonalized by generalized
polariton transformation. Resulting dispersion relations
coincide with the classical ones obtained by the solution of
wave equation, the corresponding mode decomposition is, however,
orthogonal and complete in the enlarged Hilbert space.
\end{abstract}


\section{Introduction}

The  investigation of electromagnetic field in  dielectrics has
attracted growing attention recently \cite{H58,KVW87,A87,G91,HB92}.
 Particularly, quantum aspects of this problem are of current
interest due to the potential applications in technology of
nanostructures.
This research includes investigation of quantum
wells embedded in  microcavities  \cite{S94} and generation and
propagation of nonclassical states of light. In this connection,
the quantization of
electromagnetic field  in inhomogenous time-spatially
dispersive linear medium  represents a nontrivial
problem \cite{KD93}.

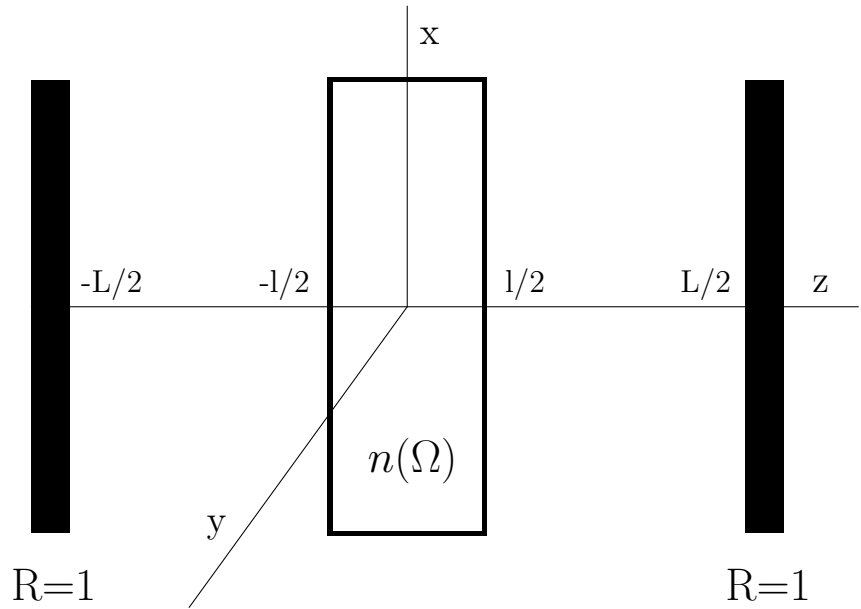
\begin{figure}

\unitlength=1.00mm
\special{em:linewidth 0.4pt}
\linethickness{1.5pt}
\begin{picture}(140.00,134.00)
\emline{20.00}{70.00}{1}{130.00}{70.00}{2}
\emline{70.00}{70.00}{3}{70.00}{110.00}{4}
\emline{70.00}{70.00}{5}{41.00}{30.00}{6}
\put(20.00,40.00){\rule{5.00\unitlength}{60.00\unitlength}}
\put(115.00,40.00){\rule{5.00\unitlength}{60.00\unitlength}}
\put(125.00,72.00){\makebox(0,0)[cb]{\large z}}
\put(109.00,72.00){\makebox(0,0)[cb]{ L/2}}
\put(85.00,72.00){\makebox(0,0)[cb]{ l/2}}
\put(53.00,72.00){\makebox(0,0)[cb]{ -l/2}}
\put(30.00,72.00){\makebox(0,0)[cb]{ -L/2}}
\put(73.00,105.00){\makebox(0,0)[cb]{\large x}}
\put(46.00,40.00){\makebox(0,0)[rb]{\large y}}
\put(60.00,40.00){\framebox(20.00,60.00)[cc]{}}
\put(70.00,48.00){\makebox(0,0)[cb]{ \Large $n(\Omega) $}}
\put(23.00,33.00){\makebox(0,0)[cc]{\Large R=1}}
\put(118.00,33.00){\makebox(0,0)[cc]{\Large R=1}}
\end{picture}

\caption{Geometry of closed cavity with inhomogeneity in the z--direction.}
\label{fig1}
\end{figure}

Our considerations are motivated by
standard electromagnetic theory \cite{Sagna}.  Suppose
for  concreteness the geometry of
closed cavity (R=1) with dispersive inhomogeneity
(refractive index $n(\Omega)$) along the z--axis
as   sketched in the  Fig. \ref{fig1}.
For the sake of simplicity
the s--polarization of electric field only will be assumed in the
folloving. Using the Maxwell equations, the eigenmodes  of
the cavity with the time dependance $e^{i\Omega t}$ may
be specified as solution
of Helmholtz (time independent) wave equation
\begin{equation}
 \biggl[\Delta  + \frac{\Omega^2}{c^2}(1+ \theta(z) \chi(\Omega))
\biggr] E = 0,
 \label{class}
\end{equation}
susceptibility being $ \chi(\Omega) = \chi' - i \chi''
= n^2(\Omega) -1.$
 The inhomogeneity is included in the characteristic function
$$\theta(z) =  \left\{ \matrix {
1  &{\rm for}\; |z| \le l/2, \cr
0  &{\rm for}\; |z| > l/2 . \cr}
\right.
$$
This equation may be simply solved in the regions where the
coefficients are  continuous function of $z.$
The electric field $E$ must be  continuous together with its
first derivation $\frac{\partial E}{\partial z}$  for each
$|z| \le L/2.$
The boundary conditions at $\pm L/2$ are given as  $E(\pm L/2) = 0$
and they  yield the dispersion  relation
\begin{eqnarray}
Q_z'  \tan\biggl[ Q_z \frac{L-l}{2} \biggr]=
Q_z \cot\biggl[ Q_z'\frac{l}{2} \biggr], \nonumber  \\
\label{disp}\\
Q_z'  \tan\biggl[ Q_z \frac{L-l}{2} \biggr]=
- Q_z \tan \biggl[ Q_z'\frac{l}{2} \biggr], \nonumber
\end{eqnarray}
where
$$  Q_z^2 =  \frac{ \Omega^2}{c^2} - q^2 \;\;\;{\rm and}\;\;\;
    Q_z'^2 =  n^2(\Omega) \frac{ \Omega^2}{c^2} - q^2.$$
Here ${\bf q}$ represents the 2D component of wave vector paralel
to the  boundaries. This  transcendent equation may be solved
yielding the discrete set of eigenvalues. Nevertheless, a simple
analysis shows that the corresponding eigenfunctions $E_{q,m}(z)$
 are not orthogonal since the ``potential"  $\chi(\Omega)$
 depends on the frequency.  This is the source of
theoretical troubles,
since the decomposition of electric field is  becoming
questionable.  The quasinormal  modes in leaky macrocavity
were used in Refs.   \cite{Le1,Le2,Le3}.

The purpose of this contribution is to clarify the quantum
meaning of this classical electromagnetic problem.
Adopting the canonical quantization scheme formulated by Huttner and Barnett
\cite{HB92}, we will formulate and solve the problem of quantization
of electromagnetic field in closed cavity with dispersive
inhomogeneity.
Normal modes are associated with (generalized) polariton transformation.
Two extreme cases of this formulation may be distinguished as
problem of  exciton confined in quantum well embedded in
microcavity \cite{S94} and  the
above mentioned classical problem  of leaky macrocavity.
Polariton solution   exactly yields the  orthogonal
decomposition.
Additional degrees of freedom are  associated with matter
excitations.  For the sake of simplicity all the quantum
considerations will be performed for special form of singular refractive
index. In Appendix  \ref{A}  this result will be  extended
to the general form of an arbitrary refractive index
fulfilling the Kramers--Kronig relations.

\section{Canonical quantization}

Let us formulate canonical  description of  interaction  of transversal
electromagnetic field with   matter.
Neglecting other losses the Lagrangian  reads
\begin{equation}
L = \int d^3 {\bf r} \;\; {\cal L}({\bf r}),
\label{start}
\end{equation}
where Lagrangian density is
$
{\cal L}  = {\cal L}_{em} + {\cal L}_{mat} + {\cal L}_{int},$
\begin{eqnarray}
\label{em}
{\cal L}_{em} =  \frac{  \epsilon_0}{2} [  \dot {\bf A}^2 -
 c^2 (\nabla \times {\bf A })^2] , \\
{\cal L}_{mat} =  \frac{\rho}{2} \; [ \dot {\bf X}^2 -
\omega_0^2 {\bf X}^2 ],
\label{mat}            \\
\label{int}
{\cal L}_{int} =  - \alpha   {\bf A \cdot  \dot X},
\end{eqnarray}
boldface characters  denote vectors  and dot  means time derivation
$\frac {\partial}{\partial t}.$
Electromagnetic part is represented by vector field {\bf A}
defined in the whole cavity.
 Polarisation part is  modeled by harmonic oscillator field with
amplitude vector {\bf X}, which is  non--zero only in the
interval of inhomogeneity   $ |z| \le l/2,$
 $\omega_0$ being the frequency and  $\rho$  density.
 Interaction
of both fields is  characterized by interaction constant $\alpha.$
For the simplicity,     linearly
polarized fields with polarization paralel to discontinuity
planes $z= \pm l/2$  will be assumed (s--polarization)  in the following.
The vector field {\bf A} may be interpreted as electric intensity
and both fields may be represented as (real) scalar fields.
The  Lagrange-Euler equations then read
\begin{eqnarray}
\epsilon_0 \biggl[ \frac{\partial^2}{\partial t^2} - c^2 \Delta
\biggr] A + \alpha \frac{\partial}{\partial t} X  = 0,\\
\rho \biggl[ \frac{\partial^2}{\partial t^2} + \omega_0^2
\biggr] X - \alpha \frac{\partial }{\partial t} A = 0
\end{eqnarray}
yielding exactly the wave equation (\ref{class}) for
susceptibility
\begin{equation}
 \chi =  \frac{\alpha^2  }{\epsilon_0  \rho
(\omega_0^2 - \Omega^2)}.
\label{susc}
\end{equation}

 \subsection{Free field  decompositions}

 The interacting fields may be quantized
using expansion in  orthogonal basis relevant to respective
free fields.
Standard approach \cite{HB92} may be used
in the plane $xy$ perpendicular to the direction of inhomogeneity,
since ${\bf q}$  is conserved due to the translation symmetry.
 In the
$z$--direction, the eigenfunction for light and matter excitation
parts should be distinguished. Assuming  $ e^{i {\bf q \tau}}/(2\pi)$
dependence, ${\bf \tau}$ being
the projection of 3D ${\bf r}$ vector into the $xy$--plane,
  $\{\varphi_m(z)\}$ are  the solutions of time--independent wave
equation
\begin{equation}
\biggl[ \frac{d^2}{dz^2}  +  Q_m^2\biggr]\varphi_m(z) = 0;
\;\;\; Q_m^2 = \frac{\Omega_m^2}{c^2} - q^2,
\label{eigen}
\end{equation}
fulfilling the given boundary conditions on the interval
$ |z| \le L/2 .$ Assuming  for concreteness
 perfect reflection on the end mirrors,
we have   $Q_m = {\pi  m }/{L}; \;\; m=1,2,3,....$
Consequently frequencies are
quantized as    $ \Omega_m(q) = c \sqrt{(\pi m /L)^2 +  q^2},\;
q = |{\bf q}|. $  Corresponding eigenfunctions are given as
   $ \varphi_m(z)=\sqrt{2} \sin\bigl[ m \pi (z/L +1/2) \bigr].$
The inhomogeneity of matter excitations is included in the definition of
eigenfunctions $\chi_{{\bf q}, \xi}({\bf r}), $  since they are
non--zero only on the interval $ |z| \le l/2.$
 Two sets of functions
 $\varphi_{{\bf q}, m}({\bf r}) $ and $\chi_{{\bf q}, \xi}({\bf r})$
 defined in the 3D space are
  orthogonal and complete  in the volumes of quantization
\begin{eqnarray}
\int d^3{\bf r}\; \varphi_{{\bf q}, m}({\bf r}) \varphi^{*}_
{{\bf q'}, n}({\bf r})=
\delta({\bf q} - {\bf q'}) \delta_{m,n}, \;\;\;
\sum_m \int d^2 {\bf q}\;
\varphi_{{\bf q}, m}({\bf r}) \varphi^*_{{\bf q}, m}({\bf r'})
= \delta({\bf r}-
{\bf r'}), \nonumber \\
\int d^3{\bf r}\; \chi_{{\bf q}, \xi}({\bf r})\chi^{*}_{{\bf q},\eta}({\bf r})
=\delta({\bf q} - {\bf q'}) \delta_{\xi,\eta} , \;\;\;
\sum_{\xi} \int d^2 {\bf q}\;
\chi_{{\bf q}, \xi}({\bf r})\chi^*_{{\bf q}, \xi}({\bf r'})
=\delta({\bf r}-
{\bf r'}) \theta(z). \nonumber
\end{eqnarray}
The  cross--products  are given by  matrix elements
\begin{eqnarray}
\int \varphi_{{\bf q}, m}({\bf r}) \chi^{*}_{{\bf q'}, \xi}({\bf r}) d^3{\bf r}
 = \delta({\bf q}- {\bf q'}) \; K_{m,\xi},
\nonumber \\
\label{ka} \\
\nonumber K_{m,\xi}   =\int_{-l/2}^{l/2}
\varphi_{m}({z}) \chi_{\xi}(z) d z , \;\; K_{m,\xi}= K_{m,\xi}^{*}.
\end{eqnarray}
Here the full set of  functions  $\{\chi_{\xi}({z})\}$ create an
  orthogonal and complete system of functions
on the interval $ |z| \le l/2 $.
The  explicit form of correct  exciton wave functions
$\chi_{{\bf q},\xi}(z) $  represents  very complex problem, solution
of which is beyond the scope of this contribution.
The basis used in decomposition of matter excitations will be
considered as given and
index $\xi$ will be
intuitively interpreted  as energy levels of free matter excitations.
In the following the wave functions of electromagnetic field will
be consistently enumerated by Latin indices, whereas Greek ones
will be used for the  matter   excitations.

Even if we started with the Lagrangian of
harmonic oscillator  field (\ref{mat}), we will incorporate into our
scheme also a little  more general models  corresponding to  the
 finite number of energy levels $\xi$
in the  decomposition of harmonic oscillator.
This technique corresponds  to quantization with
confinements tending to spatial dispersion.
 Particularly, the matter excitations in  the
ground state only   correspond to the case of single exciton
 confined  in quantum well \cite{S94}.
For this purpose, the range of sumation in respective
decompositions of matter excitations  will  not be specified
explicitly and will be mentioned  in the
  discussion of the final results only.
The respective expansions used in the following then read
\begin{equation}
A({\bf r},t) = \frac{1}{2\pi}
\sum_{m} \int d^2 {\bf q}\;  A_{{\bf q}, m}(t)
 \;\varphi_{{\bf q}, m}({\bf r}),
\end{equation}
\begin{equation}
X({\bf r},t) =\frac{1}{2\pi}
 \sum_{\xi} \int d^2 {\bf q} \; X_{{\bf q}, \xi}(t) \; \chi_{{\bf q},
\xi}({\bf r}).
\end{equation}

\subsection{Hamiltonian formalism}

The total Lagrangian
${ L}  =  { L}_{em} + { L}_{mat} + { L}_{int} $
   may be  quantized as
\begin{eqnarray}
{ L}_{em} = \epsilon_0 \sum_{m} \int' d^2 {\bf q} \biggl[
|\dot A_{{\bf q}, m}|^2   - \Omega_m^2(q) | A_{{\bf q}, m}|^2\biggr] ,
\\
{ L}_{mat} = \rho  \sum_{\xi} \int' d^2 {\bf q} \biggl[
|\dot X_{{\bf q}, \xi}|^2   - \omega_0^2 | X_{{\bf q}, \xi}|^2\biggr],
\\
{ L}_{int} = - {\alpha }   \sum_{m,\xi}  \int' d^2 {\bf q} \; K_{m,\xi}
\biggl[ A_{{\bf q}, m} \cdot \dot  X^{*}_{{\bf q}, \xi}  +c.c.
\biggr].
\end{eqnarray}
Here the   prime  means the integration over the half of the reciprocal space.
Canonically conjugated variables  are given as
\begin{equation}
A_{{\bf q}, m}^{*} \rightarrow P_{{\bf q, m}} = \frac{\partial { L}}
{\partial \dot A^{*}_{{\bf q}, m}} =  \epsilon_0 \dot A_{{\bf q}, m},
\end{equation}
\begin{equation}
X_{{\bf q}, \xi}^{*} \rightarrow Y_{{\bf q}, \xi} = \frac{\partial { L}}
{\partial \dot X^{*}_{{\bf q}, \xi}} =  \rho  \dot X_{{\bf q}, \xi} -
\alpha   \sum_n  K_{n,\xi}  A_{{\bf q}, n},
\end{equation}
and as  the complex conjugated relations.  Hamiltonian reads
\begin{equation}
 H= \epsilon_0 \sum_{m} \int' d^2 {\bf q} \; \biggl[ |\dot A_{{\bf q}, m}|^2  +
 \Omega_m^2(q) |A_{{\bf q}, m}|^2 \biggr] +
 \rho \sum_{\xi} \int' d^2 {\bf q} \; \biggl[ |\dot X_{{\bf q}, \xi}|^2  +
 \omega_0^2 |X_{{\bf q}, \xi}|^2 \biggr].
\end{equation}
Standard quantization is prescribed by commutation relations between operators
\begin{equation}
\biggl[ A_{{\bf q}, m}, P_{{\bf q'}, n}^{*} \biggr] = {i\hbar}
\delta_{m,n}\; \delta({\bf q}-{\bf q'}),
\end{equation}
\begin{equation}
\biggl[  X_{{\bf q}, \xi}, Y_{{\bf q'}, \eta}^{*}\biggr]
= {i\hbar} \delta_{\xi,\eta} \;\delta({\bf q}-{\bf q'}).
\end{equation}
Annihilation operators of electromagnetic field
\begin{equation}
a_{{\bf q}, m} = \sqrt{\frac{\epsilon_0}{2\hbar\Omega_m( q)}}
\biggl[ \Omega_m(q) A_{{\bf q}, m} + \frac{i}{\epsilon_0}
P_{{\bf q}, m}\biggr]
\end{equation}
and matter excitations
\begin{equation}
b_{{\bf q}, \xi} = \sqrt{\frac{\rho}{2\hbar\omega_0}}
\biggl[ \omega_0 X_{{\bf q}, \xi} + \frac{i}{\rho}  Y_{{\bf q}, \xi}\biggr]
\end{equation}
are fulfilling the ordinary boson commutation relations
\begin{equation}
\biggl[a_{{\bf q}, m},  a_{{\bf q'}, n}^{\dagger}\biggr] =
\delta_{m,n} \delta( {\bf q}- {\bf q'}),
\end{equation}
\begin{equation}
\biggl[b_{{\bf q}, \xi},  b_{{\bf q'}, \eta}^{\dagger}\biggr] =
\delta_{\xi,\eta} \delta( {\bf q}-
{\bf q'}).
\end{equation}
The definition may be extended to full space of ${\bf q}$ vectors
as
\begin{equation}
a_{{-\bf q}, m} = \sqrt{\frac{\epsilon_0}{2\hbar\Omega_m( q)}}
\biggl[ \Omega_m(q) A_{{\bf q}, m}^{*} + \frac{i}{\epsilon_0}
P_{{\bf q}, m}^{*}\biggr],
\end{equation}
and
\begin{equation}
b_{{- \bf q}, \xi} = \sqrt{\frac{\rho}{2\hbar\omega_0}}
\biggl[ \omega_0 X_{{\bf q}, \xi}^{*} + \frac{i}{\rho}
Y_{{\bf q}, \xi}^{*}\biggr].
\end{equation}
Hamiltonian then reads
\begin{eqnarray}
 H &=& \sum_m \int d^2 {\bf q}\;\; \hbar \Omega_m(q) \;
a_{{\bf q}, m}^{\dagger} a_{{\bf q}, m}  +
 \sum_{\xi} \int d^2 {\bf q}\;\ \hbar \omega_0 \;\;
b_{{\bf q}, \xi}^{\dagger} b_{{\bf q}, \xi} +
\nonumber \\
&+&  \frac{i\hbar}{2} G \sqrt{\omega_0 }
 \sum_{n,\xi} \int d^2 {\bf q} \;\; \frac {K_{n,\xi}}{\sqrt{\Omega_n(q)}}
 (a_{{\bf q},n} + a_{{-\bf q},n}^{\dagger}) \cdot
(b_{{\bf q},\xi}^{\dagger} -  b_{{-\bf q},\xi})
\label{ham} \\
&+& \nonumber \frac{\hbar}{4}  G^2\; \sum_{n,m} \int d^2 {\bf q}
\frac{D_{n,m}}{\sqrt{\Omega_m(q)\Omega_n(q)}}
(a_{{\bf q},n} + a_{{-\bf q},n}^{\dagger}) \cdot
 (a_{{\bf q},m} + a_{{-\bf q},m}^{\dagger}),
\end{eqnarray}
 where
\begin{equation}
 D_{n,m} = \sum_{\xi}  K_{n,\xi} K_{m,\xi}
\label{de}
\end{equation}
 and  the  effective interaction constant is abbreviated as
 \begin{equation}
  G = {\alpha}/{ \sqrt{\epsilon_0 \rho}}.
\label{G}
\end{equation}

The Hamiltonian of this type has already been  investigated
recently in connection with polariton effects \cite{S94}
and exactly corresponds to the
many--exciton lines interacting with eletromagnetic field
   discussed in  Ref. \cite{Sav94}.

\section{Dispersion relations}

 General form of polariton transformation
 diagonalizing the Hamiltonian is given as
\begin{equation}
B_{{\bf q},\Omega} = \sum_m W_{{\bf q},m} a_{{\bf q},m} +
\sum_{\xi} X_{{\bf q},\xi} b_{{\bf q},\xi} +
\sum_m Y_{{\bf q},m} a_{-{\bf q},m}^{\dagger} +
\sum_{\xi} Z_{{\bf q},\xi} b_{-{\bf q},\xi}^{\dagger}.
\label{tran}
\end{equation}
 Index $m$ exhausts all the cavity modes and $\xi$ similarly does
all the modes  of decomposition of matter excitations.
Standard diagonalization  condition
\begin{equation}
 \biggl[  B_{{\bf q},\Omega} , H \biggr] = \hbar \Omega B_{{\bf q},\Omega}
\end{equation}
yields the dispersion relation for eigenfrequency
$\Omega$ and relations for coefficients in (\ref{tran}).
The anticipated operator solution  is normalized with
respect  to the boson  commutation relation
\begin{equation}
[ B_{{\bf q},\Omega}, B^{\dagger}_{{\bf q'},\Omega'}]  =
\delta({\bf q}- {\bf q'}) \delta_{\Omega, \Omega'}.
\end{equation}
Straightforward but lengthy  calculations  lead to the following
  equations (dependence on ${\bf q}$ will be omitted for brevity)
\begin{eqnarray}
(\Omega_m - \Omega) W_m + \frac{1}{2} G^2 \sum_k
\frac{D_{m,k}}{\sqrt{\Omega_m\Omega_k}}
(W_k - Y_k)  + \frac{i}{2} G \frac{\sqrt{\omega_0}} {\sqrt{\Omega_m}}
\sum_{\xi}  K_{m,\xi} (X_{\xi} + Z_{\xi}) =  0,\\ \nonumber
\\
(\Omega_m + \Omega) Y_m - \frac{1}{2} G^2 \sum_k
\frac{D_{m,k}}{\sqrt{\Omega_m\Omega_k}}
(W_k - Y_k)  - \frac{i}{2} G \frac{\sqrt{\omega_0}} {\sqrt{\Omega_m}}
\sum_{\xi}  K_{m,\xi} (X_{\xi} + Z_{\xi}) =  0,\\ \nonumber
\\
(\Omega - \omega_0) X_{\xi}   + \frac{i}{2} G  \sqrt{\omega_0}
\sum_{m}
\frac{K_{m,\xi}}{\sqrt{\Omega_m}} (W_m - Y_m) =  0,\\ \nonumber
\\
(\Omega + \omega_0) Z_{\xi}   - \frac{i}{2} G  \sqrt{\omega_0}
\sum_{m}
\frac{K_{m,\xi}}{\sqrt{\Omega_m}} (W_m - Y_m) =  0.
\end{eqnarray}
Finally, the dispersion relation follows as the condition for existence of
non--trivial solution of linear equations for $T_m$
\begin{equation}
T_m = \frac{G^2 \Omega^2} {(\omega_0^2 - \Omega^2)
(\Omega_m^2 - \Omega^2)} \sum_n D_{m,n} T_n ,
\label{syst}
\end{equation}
where $ T_m =  (W_m - Y_m)/ \sqrt{\Omega_m}. $

 The algebraic treatment may be equivalently replaced by an
analytical method.
The recurent  system of linear equations
(\ref{syst}) may be rewritten to the form of  integro--differential
equation.  Hence the
quantum problem is  related to the classical solution
of wave equation with dispersive inhomogeneous medium.
Let us define formally the function
$$A({\bf r})=\sum_m \int d^2 {\bf q}\; T_m \; \varphi_{{\bf q},m}({\bf
r}),$$
which is continuous and has continuous derivation $ dA/dz$ on the
interval $ |z| \le L/2.$
After simple manipulations using relations (\ref{eigen}),
(\ref{ka}) and (\ref{de}) we find that it fulfills the equation
\begin{equation}
\biggl[ \frac{d^2}{dz^2}  + {Q_z^2} \biggr]\; A(z) =
- \theta(z) \frac{G^2 \Omega^2 }{c^2(\omega_0^2 -
\Omega^2)} \int_{-l/2}^{l/2}
\sum_{\xi} \chi_{\xi}(z) \chi_{\xi}(z') A(z')\; dz'\;,
\label{dif}
\end{equation}
where $  Q_z^2 =  (\Omega/c)^2 - q^2.$
This alternative representation of  (\ref{syst})
may be interpreted as scalar wave equation with the
time--spatially dispersive  inhomogeneity along the $z$--axis.
Consequently, we are able to associate the quantum problem--diagonalization
of the Hamiltonian (\ref{ham}), with the classical solution  in
electromagnetic theory.  This analogy represents a powerful tool
for specification of dispersion relations \cite{Tass90}.
The wave equation
(\ref{dif})  may be solved separately  in the regions where all the
functions are continuous.
Dispersion relation is then given as necessary condition for
continuity of the solution and its first derivation on the
boundary  $|z| = l/2.$
The general solution of eq. (\ref{dif}) on
the interval
$ |z| \le l/2 $ is given
as superposition of particular and fundamental  solutions
\begin{equation}
A(z) =  \sum_{\xi} c_{\xi} \int dz' G(z,z') \chi_{\xi}(z') +
A_2 \; e^{iQ_z z} + B_2 \; e^{-iQ_zz} ,
\end{equation}
  $A_2$ and  $B_2$ being  general multiplicators in  fundamental
solution, $ G(z,z')$  being the   Green function of the operator
$ d^2/dz^2 +  Q_z^2. $ We may take  the explicit  form
$$ G(z,z') =   -\frac{1}{2 Q_z} \sin(Q_z|z-z'|). $$
  Coefficients  $c_{\xi}$  of particular solution
are given in accordance with  the relations
\begin{eqnarray}
\nonumber
c_{\xi} \equiv  - \frac{G^2 \Omega^2 }{c^2(\omega_0^2 - \Omega^2)}
\int dz  \chi_{\xi}(z) A(z) =
- \frac{G^2 \Omega^2 }{c^2(\omega_0^2 - \Omega^2)}
\biggl\{ \sum_{\eta} c_{\eta} \int dz\; dz'
\chi_{\xi}(z) G(z,z') \chi_{\eta}(z')
\\ \label{ce}  \\ \nonumber
+A_2  \int dz'\;\chi_{\xi}(z') e^{iQ_z z'} +
B_2 \int dz'\;\chi_{\xi}(z') e^{-iQ_z z'}
\biggr\}.
\end{eqnarray}
  Parametrizing the solutions on the intervals
$  L/2 \le |z| \le  l/2 $   as
$ A_j e^{iQz} +  B_j e^{-iQz}; \;\; j=1,3$ the dispersion relation
may be   found in closed form for limited number of terms   $\xi$
 for any mode decomposition $\chi_{\xi}(z).$

Algebraic method will be used in the following sections
\ref{one}
 and \ref{many}, whereas the analytical  method will
be applied on the analogy of classical problem in the section
\ref{cl}.

\subsection{One--exciton  dispersion relation}
\label{one}

Closed form of dispersion relation may be  find easily
 in some  special   cases.  Particularly, exciton  confined
in  quantum well (QW)  is  characterized by the only
energy level in the expansion of matter excitations $\xi = 0$
(ground state of excitations).
The integro--differential equation  (\ref{dif}) indicates the
spatial dispersion. Nevertheless, the
dispersion relation may  be   found without solving it,
 since the coefficients
$D_{m,k}$  are  factorized  as product
$D_{n,m} = K_{n,0} \cdot K_{m,0}.$
Equations (\ref{syst}) then directly yield the  necessary condition
\begin{equation}
 \frac{G^2 \Omega^2 }{\omega_0^2 - \Omega^2}
\sum_n \frac{K_{m,0}^2}{\Omega_m^2 - \Omega^2}  =  1,
\end{equation}
representing the dispersion relation of
QW polaritons embedded in microcavity
\cite{S94}.

\subsection{Many--exciton dispersion  relations}
\label{many}

Dispersion relation for  many excitons interacting with
electromagnetic field will be demonstrated on the
 explicit example of two  excitons  ($\xi = 0, 1$).
Taking into account the form of the kernel $D_{m,n}=
K_{m,0}K_{n,0} + K_{m,1}K_{n,1},$
the system of linear equations (\ref{syst}) may be rewritten as
\begin{eqnarray}
x_0 = x_0  \frac{G^2 \Omega^2 }{\omega_0^2 - \Omega^2}
\sum_m \frac{K_{m,0}^2 }{\Omega_m^2 - \Omega^2} +
      x_1  \frac{G^2 \Omega^2 }{\omega_0^2 - \Omega^2}
\sum_m \frac{K_{m,0}K_{m,1} }{\Omega_m^2 - \Omega^2} \nonumber
\\
\\
x_1 = x_0  \frac{G^2 \Omega^2 }{\omega_0^2 - \Omega^2}
\sum_m \frac{K_{m,0}K_{m,1} }{\Omega_m^2 - \Omega^2} +
      x_1  \frac{G^2 \Omega^2 }{\omega_0^2 - \Omega^2}
\sum_m \frac{K_{m,1}^2 }{\Omega_m^2 - \Omega^2}, \nonumber
\end{eqnarray}
where $x_{\xi} = \sum_m T_m K_{m,\xi}; \xi = 0,1.$
Dispersion relation for two exciton then reads
\begin{equation}
\biggl[
1 - \frac{G^2 \Omega^2 }{\omega_0^2 - \Omega^2}
\sum_m \frac{K_{m,0}^2 }{\Omega_m^2 - \Omega^2} \biggr]
\biggl[
1 - \frac{G^2 \Omega^2 }{\omega_0^2 - \Omega^2}
\sum_m \frac{K_{m,1}^2 }{\Omega_m^2 - \Omega^2} \biggr]
=
\biggl[  \frac{G^2 \Omega^2 }{\omega_0^2 - \Omega^2}
\sum_m \frac{K_{m,0}K_{m,1} }{\Omega_m^2 - \Omega^2}
\biggr]^2.
\end{equation}
The sums involved in the dispersion relation may further be
expressed in the analytical form as done in Ref. \cite{S94},
but this is beyond the scope of this paper. Let us only note
that the application of  analytical method yields directly the
closed form of dispersion relations.

\subsection{Classical dispersion relations}
\label{cl}

As the last special example,  the dispersion relations resulting
from the  classical electrodynamics in  dispersive
inhomogeneity (\ref{class}) will be considered from the quantum
viewpoint.
This case is characterized  by the   decomposition
of  matter excitations, which is   complete on the interval of
inhomogenity (i.e. all the  energy  lines of excitations are
included).  Relation (\ref{de}) then  reads
\begin{equation}
D_{n,m} = \sum_{\xi} \int_{-l/2}^{l/2} \varphi_{n}({z})
 \chi_{\xi}(z) d z \cdot \int_{-l/2}^{l/2} \varphi_{m}({z'})
 \chi_{\xi}(z') d z' = \int_{-l/2}^{l/2} \varphi_{n}({z})
 \varphi_{m}(z) d z .
\label{specde}
\end{equation}
Spatial dispersion in  (\ref{dif})
 disappears  yielding scalar wave
 equation  identical with Lagrange--Euler
equation (\ref{class})  for the susceptibility (\ref{susc})
and for the dispersion relation (\ref{disp}).
Consequently,
the classical  solution yielding   non--orthogonal eigenmode
functions was completed by  the  fully quantum  treatment characterized
by the same dispersion relations but   orthogonal decomposition.

\section{Conclusion}

The problem of canonical quantization of electromagnetic field  in
 linear  dispersive
inhomogeneous media was formulated in  terms of overlaping of
wave functions related to quantization of free  electromagnetic  and
matter excitation fields.
 Diagonalization  is given  by generalized
polariton (Hopfield) transformation.  Within the  classical
  electrodynamics, it
may be also interpreted as  the  wave in spatially  dispersive
medium.
The description of macroscopic
inhomogeneity and quantum well polaritons may be unified in this
way into the same framework.

Even if quantum and classical problems yield the same dispersion relations,
there is a difference between both the treatments. Since the
electromagnetic  field itself is not conserved,
 the respective eigenfunctions are not orthogonal
representing the system of    quasinormal modes
\cite{Le1,Le2,Le3}. They
may be used for description of electromagnetic field inside the cavity,
however since completeness  and orthogonality
relations should be redefined, the description is
more complicated than in the ordinary  case of orthogonal modes.
On the other hand the quantum solution is characterized by
an ordinary orthogonal  diagonalization in the form of
 generalized polariton transformation acting
on the space of coupled electromagnetic and
matter excitation fields.
Standard description  of the time evolution may be
used, since the normal modes  are orthogonal and complete.

\section*{Acknowledgment}
I am grateful for valuable comments to P. Schwendimann, A.
Quattropani,
 V. Savona and A. Luk\v{s}. Partial  support of Swiss National Optics
Program   and internal grant of Palacky University is
acknowledged.

\appendix
\section{Inhomogeneity with an arbitrary refractive index}
\label {A}

The theory developed above describes the
lossless dispersive inhomogeneity
characterized by the real part of the  susceptibility (\ref{susc})
\begin{equation}
\chi'(\Omega)  =   \frac{G^2}{\omega_0^2 - \Omega^2}.
\label{su}
\end{equation}
Principle of superposition  may be used now to get an arbitrary
 susceptibility as done in Ref. \cite{HBE92}.
The electromagnetic field will interact with
an ansamble of independent oscillators with Lagrangian densities  as in
 (\ref{mat},\ref{int}),
 distinguished by frequencies $\omega_0, \omega_1,....$ etc. and
by
different parameters $\alpha_0, \alpha_1,$ etc.  and
$ \rho_0, \rho_1,$ etc..
In the limit of continuous frequency distribution $\omega,$ the
parameters are considered as  frequency dependent
$\alpha(\omega), \rho(\omega) $  yielding  the total contribution
to the Lagrangian  density
\begin{eqnarray}
{\cal L}_{mat} = \int_0^{\infty}  d\omega \frac{\rho(\omega)}{2} \;
[ \dot {\bf X}_{\omega}^2 - \omega^2 {\bf X}_{\omega}^2 ], \\
{\cal L}_{int} =  - \int_0^{\infty} d\omega \alpha(\omega)
 {\bf A \cdot  \dot X}_{\omega}.
\end{eqnarray}
Further development of the theory runs  similarly as in  the
single--frequency case: Matter operators are denoted by
modal index $\xi$ and by an additional continuous index $\omega.$
The effective interaction parameter (\ref{G}) is
freqency dependent  $G(\omega).$  The final system of linear
equations (\ref{syst}) reads
\begin{equation}
T_m =  \frac{ \Omega^2 }{\Omega_m^2 - \Omega^2}
 \int_0^{\infty}  \frac{G^2(\omega) }{\omega^2 - \Omega^2 }\;
 d\omega \; \sum_n D_{m,n} T_n,
\end{equation}
what corresponds to the
real   part of the  susceptibility
\begin{equation}
\chi'(\Omega) =  \int_{0}^{\infty}  d\omega
 \frac{G^2(\omega)   }{\omega^2 - \Omega^2}.
\label{chi'}
\end{equation}

Due to the Kramers--Kronig relations \cite{HBE92}, the real and
imaginary part of susceptibility are mutually
related by Hilbert transformation as
\begin{eqnarray}
\chi'(\Omega) = \frac{2}{\pi} \int_{0}^{\infty} \frac{\omega \chi''(\omega)}
{\omega^2 - \Omega^2} d \omega, \\
\chi''(\Omega) = -  \frac{2 \Omega}{\pi} \int_{0}^{\infty} \frac{
\chi'(\omega)}
{\omega^2 - \Omega^2} d \omega.
\end{eqnarray}
The imaginary part of lossless and general cases  therefore read
\begin{equation}
\chi''(\Omega)  =   -  {\pi G^2}
\delta(\Omega^2 - \omega_0^2)
\end{equation}
and
\begin{equation}
\chi''(\Omega)  = -  {\pi}   \int_{0}^{\infty} d \omega
 {G^2(\omega)} \delta(\omega^2 - \Omega^2),
\label{chi''}
\end{equation}
respectively.
The relations (\ref{chi'}) and  (\ref{chi''})
 represent the desired extension of the theory with
single--frequency of matter excitations and  singular
susceptibility into an arbitrary case characterized by general
refractive index (susceptibility).

\begin{thebibliography}{9999}



\bibitem{H58}
J. J. Hopfield,
\newblock Phys. Rev.  {\bf  112}, (1958) 1555.

\bibitem{KVW87}
L. Kn\"{o}ll, W. Vogel, D.--G. Welsh,
\newblock Phys. Rev. A {\bf  36}, (1987) 3803.

\bibitem{A87}
I. Abram,
\newblock Phys. Rev. A{\bf  35},  (1987) 4661.


\bibitem{G91}
R. J. Glauber and M. Lewenstein,
\newblock Phys. Rev. A {\bf  43}, (1991)  467.

\bibitem{HB92}
B. Huttner and S. M. Barnett,
\newblock Phys. Rev. A {\bf  46},  (1992) 4306.


\bibitem{S94}
V. Savona, Z. Hradil, A. Quattropani and  P. Schwendimann,
\newblock Phys. Rev. B {\bf 49}, (1994)  8774.

\bibitem{KD93}
A. N. Kireev and M. A. Dupertuis,
\newblock Mod. Phys. Letters B {\bf 7}, (1993) 1633.


\bibitem{Sagna}
P. M. N. Sagna,
\newblock  diploma thesis, EPF   Lausanne 1990, (unpublished).


\bibitem{Le1}
P. T. Leung, S. Y. Liu and K. Young,
\newblock
Phys. Rev. A {\bf  49},  (1994) 3057.


\bibitem{Le2}
P. T. Leung, S. Y. Liu and K. Young,
\newblock
Phys. Rev. A {\bf  49},  (1994) 3068.


\bibitem{Le3}
P. T. Leung, S. Y. Liu and K. Young,
\newblock
Phys. Rev. A {\bf  49},  (1994) 3982.

\bibitem{Sav94}
R. Girlanda, S. Savasta, V. Savona, A. Quattropani, and P. Schwendimann,
\newblock  { \em in  Proceedings  of the 22nd International
Conference on the Physics of Semiconductors}, ed. by D. J. Lockwood,
(World Scientific, 1995)  1396.


\bibitem{Tass90}
F. Tassone, F. Bassani and L. C. Andreani,
\newblock  Il Nuovo Cimento {\bf 12 D},  (1990)  1673.



\bibitem{HBE92}
B. Huttner and S. M. Barnett,
\newblock Europhys. Lett {\bf  18},  (1992) 487.


\end{thebibliography}
\end{document}